# Possible Room Temperature Ferromagnetism in Hydrogenated Carbon Nanotubes

Adam L. Friedman[1,3], Hyunkyung Chun[2], Yung Joon Jung[2], Don Heiman[1], Evan R. Glaser[3], and Latika Menon[1]

[1]*Department of Physics, Northeastern University, Boston, MA 02115 USA*

[2]*Department of Mechanical and Industrial Engineering, Northeastern University, Boston, MA 02115 USA*

[3]*Naval Research Laboratory, Washington, D.C. 20375 USA*

**Abstract**

We find that ferromagnetism can be induced in carbon nanotubes (CNTs) by introducing hydrogen. Multiwalled CNTs grown inside porous alumina templates contain a large density of defects resulting in significant hydrogen uptake when annealed at high temperatures. This hydrogen incorporation produces H-complex and adatom magnetism which generates a sizeable ferromagnetic moment and a Curie temperature near $T_C=1000$ K. We studied the conditions for the incorporation of hydrogen, the temperature-dependent magnetic behavior, and the dependence of the ferromagnetism on the size of the nanotubes.

Inducing magnetism in carbon structures could have far reaching consequences for future carbon-based electronics. Magnetism can be induced in carbon structures by defects such as vacancies or impurities. In the absence of defects, all of the bonding electrons are paired in π bonds. However, defects which delocalize one of the pair bonds induce excess spin polarization. This can lead to ferromagnetism when the defects are sufficiently dense. In graphite, impurity-induced ferromagnetism has been demonstrated and has been studied theoretically [1-9]. The two primary contributors to this magnetism are *H-complex* and *adatom* magnetism. In *H-vacancy complex magnetism*, hydrogen bonds to a carbon atom near a carbon vacancy creating a new σ bond, thereby inducing spin polarization in an unpaired π-electron [2,3]. The increased interlayer separation in ordered graphite provides an appropriate length scale for hydrogen adsorption and the formation of carbon interstitials and Frenkel pairs. On the other hand, a defect caused by a carbon adatom would have the same effect as in the H-complex of delocalizing the nearby π-electrons by becoming saturated by hydrogen, resulting in *adatom ferromagnetism*. However, at high defect densities a significant number of adatoms will likely form covalent bonds in nearby vacancies [2,3]. It was also postulated that oxygen and nitrogen incorporation can lead to ferromagnetism in carbon structures [5]. Following these predictions, defect-induced magnetism was found in experiments on graphite [10] and films of highly-ordered pyrolytic graphite (HOPG) [11]. Nevertheless, ferromagnetism arising from hydrogen incorporation has not been observed experimentally in CNTs because most CNT systems are well-ordered and lack a sufficient concentration of defects for hydrogen uptake. In this letter, we report on the observation of ferromagnetism in H-annealed, multiwalled CNTs. The nanotubes fabricated inside nanoporous alumina templates have thick, highly disordered walls which are ideal for hydrogen incorporation.

Multiwalled CNT were synthesized by chemical vapor deposition (CVD) in the nanopores of alumina templates. The porous alumina templates are synthesized by DC anodization of aluminum foil (99.997%) to act as a deposition substrate for the CNTs [17]. Porous alumina templates are convenient substrates for CNT synthesis because the dimensions of the pores determine the dimensions (outer diameter and length) of the CNT. CNTs are synthesized in the templates by means of a CVD process with acetylene acting as the precursor gas at 660 ºC without catalyst. The inner diameter of the CNTs is controlled by the time of the CVD process. Nanotubes were synthesized with 4 different inner diameters using CVD times of 60 minutes, 75 minutes, 90 minutes, and 100 minutes. The nanotubes were removed from the templates by dissolving the alumina then dispersed onto Si/SiO$_2$ substrates. Transmission electron microscopy (TEM) images show that the tubes are polycrystalline and highly disordered. The samples were annealed in a tube furnace containing equal parts Ar and H$_2$ up to 800 ºC. For comparison, samples were also annealed in Ar gas alone. Magnetization measurements were carried out in a Quantum Designs MPMS SQUID magnetometer and the magnetic field is applied parallel to the substrate containing the randomly-oriented tubes.

In order to study the effects of hydrogen incorporation on the CNTs we prepared samples which were, (*i*) annealed with hydrogen and argon, (*ii*) annealed with argon alone, and (*iii*) compared to an empty substrate. Figure 1(a) shows the magnetic moment as a function of applied field, $m(H)$, for the samples measured at room temperature. There is a sharp contrast between the $m(H)$ behavior for the H-annealed CNT samples when compared with the Ar-annealed sample and the empty substrate. The H-annealed CNTs exhibit ferromagnetic hysteresis with a saturation magnetic moment $m_S \sim 30\mu$emu, which is more than an order of magnitude larger than the moment measured for the other samples. Indeed, ferromagnetic impurity contamination is a major issue in such experiments [7,12], however, the large excess in $m(H)$ behavior clearly indicates that the CNT ferromagnetism is due to the hydrogen incorporation process. There is however a small paramagnetism in both the Ar-annealed samples and the substrate as seen in Fig. 1(a), which is attributed to impurities in the substrate and a paramagnetic contribution from the graphitic structure of the CNTs. It should also be emphasized here that the samples were annealed in the upstream of the steady gas flow, where any impurities emanating from the annealing quartz tube would not deposit in the sample (such impurities typically deposit in the downstream, or are swept away by the vacuum pump). Empty $Si/SiO_2$ substrates treated with hydrogen in the same fashion displayed behavior equivalent to the Ar-treated empty substrate, indicating impurities were not introduced in the annealing process. The inset in Fig. 1(a) shows the $m(H)$ behavior for the H-annealed CNT samples at T = 4 K, from which the coercive field is estimated to be ~100 Oe.

Figure 1(b) shows the dependence of magnetic moment as a function of temperature, $m(T)$. The lower data is the moment using as small applied field of H=100 Oe. The upper data is the saturation moment, $m_S(T)$, obtained after subtracting the diamagnetism. It is clear that the moments have a slow variation as a function of temperature and persist well above room temperature. The data can be adequately fit to a mean-field model, $m(T)=m_o[1-(T/T_C)^{1/2}]$. Fitting the data yields transition temperatures of 910 K and 1120 K for the low-field and saturation moments, respectively. Thus, the Curie temperature is estimated to be $T_C=1000\pm100$ K. In previous studies of ferromagnetic carbon [6,11,13], the temperature dependence was fit using a two-dimensional (2D) model that accounts for anisotropic spin waves in conjunction with the Ising model. In the present case, however, the mean-field model provides an adequate fit.

Further information on the ferromagnetism was obtained by measuring the dependence of magnetization on the annealing temperature and on the number of CNT walls. Figure 2 shows the dependence of saturation magnetization as a function of annealing temperature for the H-annealed CNTs. Note that there is no appreciable ferromagnetic moment until the annealing temperature reaches 500 °C, after which the magnetization achieves a maximum value for temperatures greater than 600 °C. The temperature necessary for making and breaking C-H bonds is estimated to be 600 °C [14,15]. For annealing temperatures just below 600 °C, there is a small ferromagnetic moment which rapidly decreases with decreasing annealing temperature well below 600 °C. This small residual magnetic moment for the low temperature annealed samples is attributed to impurities in the Si/SiO$_2$ substrate material, rather than due to the magnetic moment of the nanotubes themselves. The much larger moment of the nanotubes annealed at temperatures ≥ 600 °C as compared to lower temperature annealed tubes is another indication that random impurities are not the source of the nanotube magnetism.

The magnetic moment was also investigated as a function of the number of CNT walls in the tubes. A series of samples was synthesized with different CNT wall thicknesses by varying the inner tube diameter while keeping the outer diameter fixed at $D_{out}$=40 nm. The number of walls in the CNTs was increased by decreasing the inner diameter of the tubes. The inset shows the variation of the saturation magnetization for inner diameters from $D_{in}$=20 to 39 nm. For very thin walls, $D_{in}$= 39 nm, there is negligible magnetization. As the inner diameter decreases, the magnetization increases rapidly. Thus, the magnetic moment increases due to a greater number of C-H bonds becoming available. This confirms that the hydrogen does not attach only to the wall surfaces, but is incorporated *throughout* the highly-disordered tube walls. A cross sectional image illustrating the wall structure is shown by the transmission electron microscope (TEM) image in the inset of Fig. 3. The high density of vacancies arising from the sizeable disorder present in our CNTs allows for the uptake of large quantities of hydrogen. The increased number of individual C-H bonds creates a higher overall moment per CNT. From the measured magnetization, it is estimated that there are approximately $1.5 \times 10^{12}$ C-H bonds/ng, or 0.03 $\mu_B$/Carbon atom, calculated using the theoretical values for moment/C-H bond [5]. This indicates that the hydrogen concentration is a *few percent of the carbon density*, and is consistent with the report of Ohldag, et al. [16] for proton irradiated HOPG.

X-ray photoelectron spectroscopy (XPS) measurements have been performed to confirm the presence of H in the CNTs. Nikitin, et al. [14], showed that small amounts of hydrogen can be detected in CNTs by measuring the C1s level using XPS. Figure 3 shows a narrow linewidth (FWHM=2.6 eV) C1s peak for the Ar-only treated CNTs, which is indicative of C-C $\pi$ bonds and the absence of defect induced C-H bonds or structural rearrangements. For H-annealed CNTs, there is a sharp decrease in the peak intensity, a shift in the peak position, and an increase in the FWHM to 2.9 eV. The decrease in peak intensity corresponds to a decrease in pure, un-hydrogenated C atoms [14] and is consistent with the formation of C-H bonds. The peak shift is also consistent with the decrease in the number of C-C $\pi$ bonds leading to a decrease in the $\pi^*$ resonance and the increase of $\sigma^*$ resonances. Finally, since the XPS signal is measured for CNT samples dispersed on a Si/SiO$_2$ wafer, the low CNT density on the wafer exposes the Si and O in the wafer to the source and detector also leading to peak broadening. Electron paramagnetic resonance (EPR) studies have been performed to further confirm the absence of ferromagnetic impurities. The measurements were obtained at 9.5 GHz on both H-annealed and Ar-annealed CNTs. EPR detected no impurities in the CNTs over a wide magnetic field range (1000 – 4000 Oe). The absence of a signal in the Ar-annealed CNTs (even though defects exist) appears physically consistent. Rather than allowing for a free electron at a vacancy site, the carbons with dangling bonds can form strained bonds with carbon atoms near the vacancy, which saturate the vacancy and give rise to no EPR signal. EPR measurements indeed confirm that there were no appreciable impurities in the CNTs, judging from the absence of resonances from transition (such as Cu or Cr) or ferromagnetic (such as Fe or Ni) metals.

We summarize our justification for the exclusion of ferromagnetic impurities as a cause of the observed behavior: (1) XPS measurements revealed the presence of hydrogen in the CNTs, which studies[13] have shown results in spin polarization and can lead to the observed ferromagnetism. (2) EPR measurements showed no resonances from ferromagnetic impurities. (3) The samples that were annealed in hydrogen showed a strong ferromagnetic signal as compared to the samples annealed in argon alone and annealed and un-annealed substrates. (4) Due to the samples being annealed in the upstream of the tube furnace, it is unlikely that significant impurities would deposit on the CNTs during annealing. Taking into account all of these considerations, there is strong evidence that the ferromagnetism originated in the hypothesized C-H bonding mechanisms.

In conclusion, we have shown that CNTs synthesized using the CVD process results in highly disordered multiwalled CNTs, which are ideal for hydrogen uptake. We observe a ferromagnetic moment at room temperature which can possibly be ascribed to a combination of H-complex and adatom magnetism. As predicted by previous theoretical studies, observed ferromagnetism commences at annealing temperatures above about 600 ºC. The CNT magnetization is found to increase for increasing wall thickness of the nanotubes, indicating that hydrogen is incorporated *throughout* the tube walls.

**Acknowledgements**

This work was supported by the NSF grant ECCS 0551468.

**References:**

[1] Y. Zhang, S. Talapatra, S. Kar, R. Vajtai, S. K. Nayak, and P. M. Ajayan, Phys. Rev. Lett. **99**, 107201 (2007); and references contained therein.

[2] Y. Ma, P.O. Lehtinen, A. S. Foster, and R. M. Nieminen, Phys. Rev. B **72**, 085451 (2005).

[3] P.O. Lehtinen, A.S. Foster, Yuchen Ma, A.V. Krasheninnikov, and R. M. Nieminen, Phys. Rev. Lett. **93**, 187202 (2004).

[4] P. Esquinazi, D. Spemann, R. Hohne, A. Setzer, K.-H. Han, and T. Butz, Phys. Rev. Lett. **91**, 227201 (2003).

[5] S. Talapatra, P.G. Ganesan, T. Kim, R. Vajtai, M. Huang, M. Shima, G. Ramanath, D. Srivastava, S.C. Deevi, and P. M. Ajayan, Phys. Rev. Lett. **95**, 097201 (2005).

[6] J. Barzola-Quiquia, P. Esquinazi, M. Rothermel, D. Spemann, T. Butz, and N. Garcia, Phys. Rev. B, **76** 161403R (2007).

[7] P. Esquinazi, J. Barzola-Quiquia, D. Spemann, M. Rothermel, H. Ohldag, N. Garcia, A. Setzer, and T. Butz, J. Mag. Mag. Mater. (to be published) (2009).

[8] O. V. Yazyev, Phys. Rev. Lett., **101**, 037203 (2008).

[9] K. Kusakabe and M. Maruyama, Phys. Rev. B, **67**, 092406 (2003).

[10] H. Ohldag, T. Tyliszczak, R. Hohne, D. Spemann, P. Esquinazi, M. Ungureanu, and T. Butz, Phys. Rev. Lett., **98**, 187204 (2007).

[11] J. Cervenka, M.I. Katsnelson, and C.F.J. Flipse, Nature Physics, 5, 840 (2009).

[12] H. Zeng, R. Skomski, L. Menon, Y. Liu, S. Bandyopadhyay, and D. J. Sellmyer, Phys. Rev. B **65**, 134426 (2002).

[13] Huihao Xia, Weifeng Li, You Song, Xinmei Yang, Xiangdong Liu, Mingwen Zhao, Yueyuan Xia, Chen Song, Tian-Wei Wang, Dezhang Zhu, Jinlong Gong, and Zhiyuan Zhu, Adv. Mater., **20**, 4679 (2008).

[14] A. Nikitin, H. Ogasawara, D. Mann, R. Denecke, Z. Zhang, H. Dai, K. Cho, and A. Nilsson, Phys. Rev. Lett., **95**, 225507 (2005).

[15] Xuesong Li, Hongwei Zhu, Lijie Ci, Cailu Xu, Zongqiang Mao, Bingquig Wei, Ji Liang, and Dehai Wu, Carbon **39**, 2077 (2001).

[16] H. Ohldag, P. Esquinazi, E. Arenholz, D. Spemann, M. Rothermel, A. Setzer, T. Butz, arXiv: 0905.4315 (2009).

[17] Adam L. Friedman, Derrick Brittain, and Latika Menon, J. Chem. Phys. **127**, 154717 (2007), and references contained therein.

**Figure Captions**

Figure 1: (a) Magnetic moment versus applied magnetic field for CNTs annealed in $H_2$/Ar, Ar, and the empty substrate, measured at T=4 K. A diamagnetic background has been subtracted. The upper inset shows the hysteresis of hydrogen annealed tubes, displaying a coercivity of ~0.015 T. (b) Magnetic moment as a function of temperature for the hydrogen annealed CNTs after subtracting the diamagnetism. The upper data represents the saturation moment and the lower data represents the low-field (100 Oe) moment. The solid curves are fits to a mean-field model, with transition temperatures of 910 K and 1120 K.

Figure 2: Magnetization of CNT versus annealing temperature, showing a rapid increase for temperatures above 500 ºC, in tubes with an outer diameter of 40 nm. The inset shows the magnetization as a function of CNT inner diameter, for a sample having tubes with an outer diameter of 40 nm.

Figure 3: XPS measurement of the C1s line for Ar treated (upper curve) and hydrogen treated (lower curve) CNTs indicating hydrogen saturation in the hydrogen treated CNTs. Inset: TEM image showing the disordered CNT walls.

Figure 1a

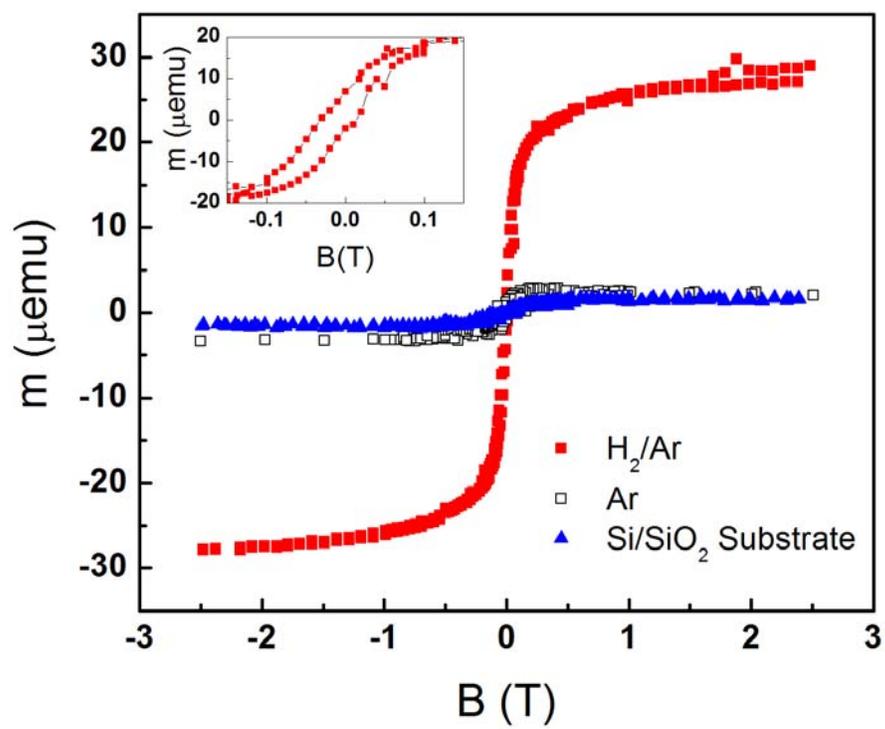

Figure 1b

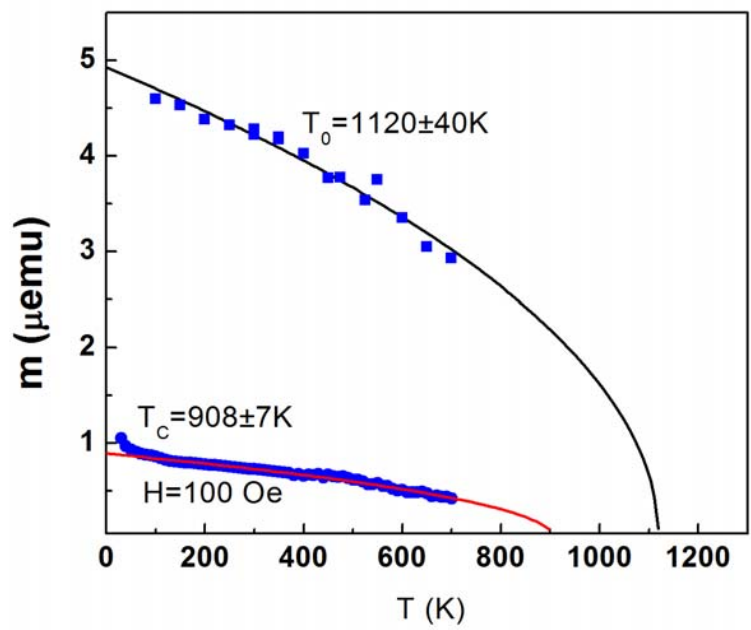

Figure 2

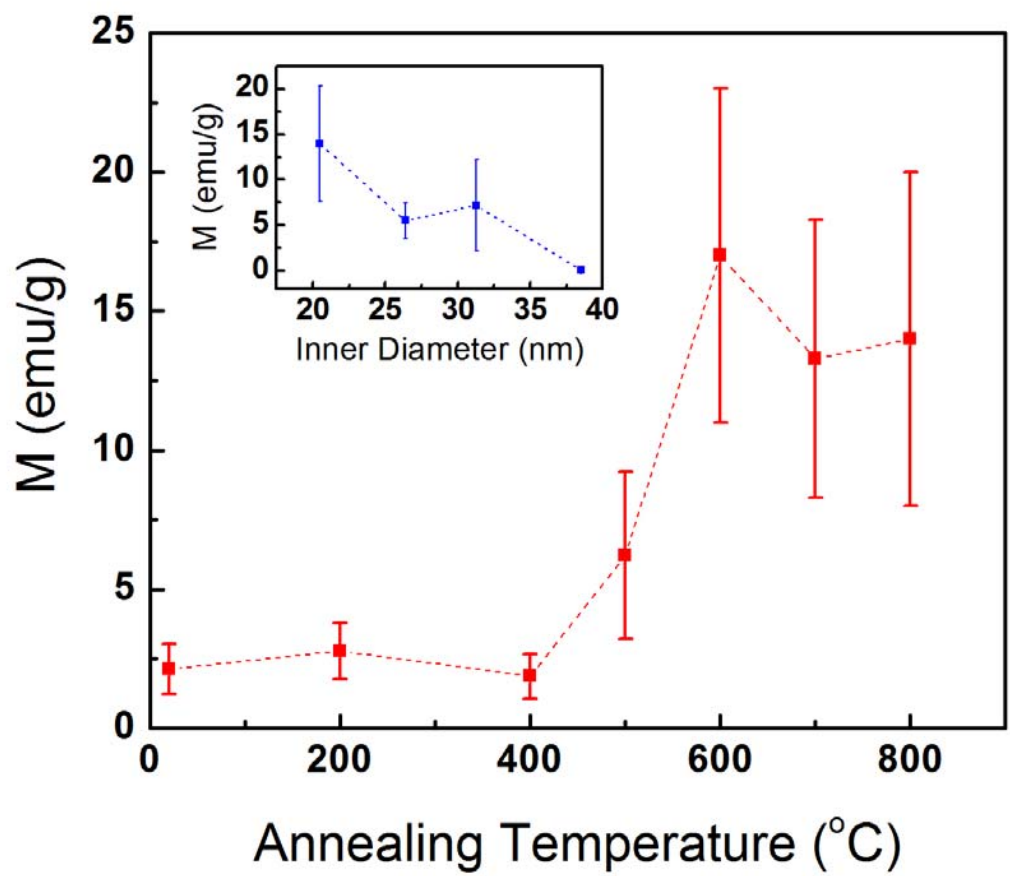

Figure 3

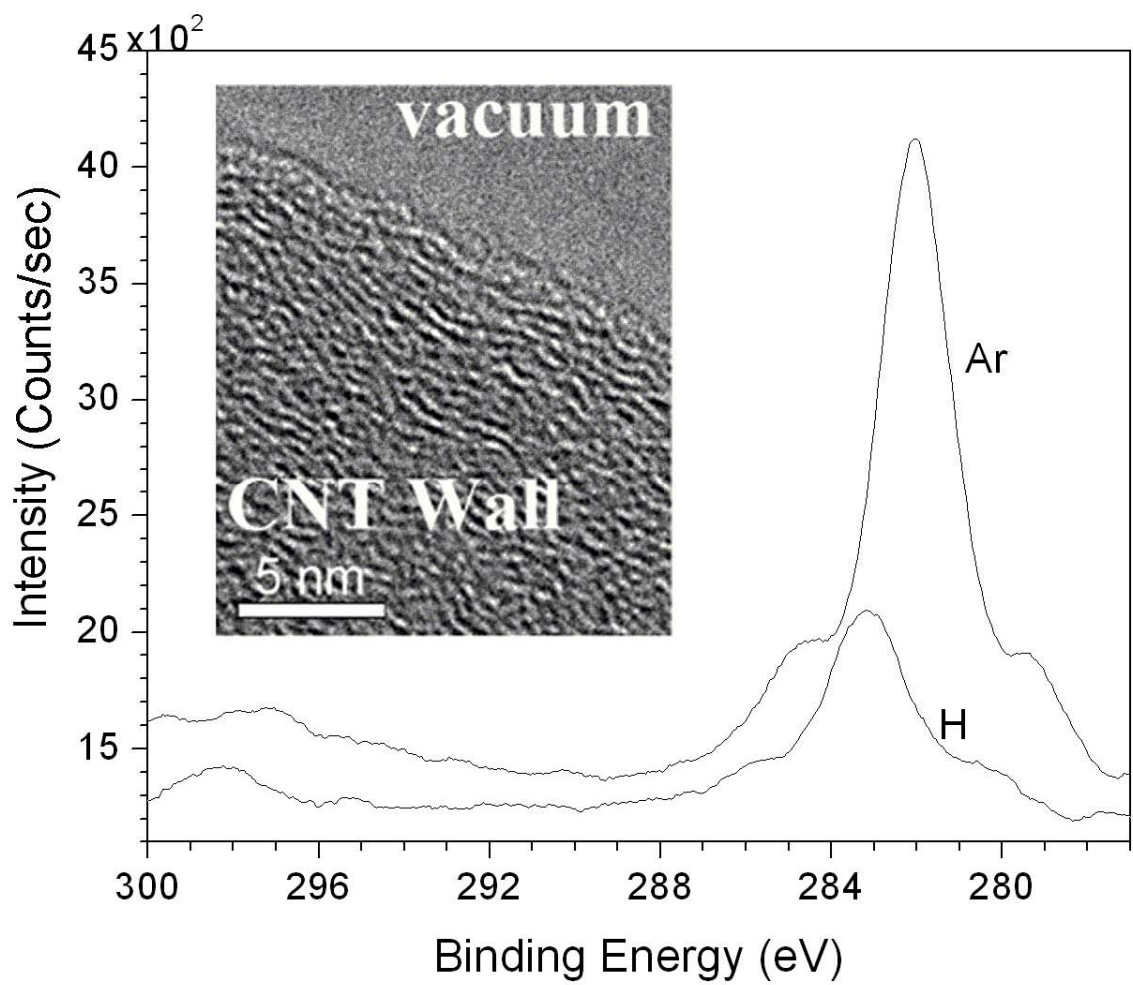